\documentclass[a4paper,11pt]{article}
\usepackage{latexsym}
\newcommand{\beq}{\begin{equation}}
\newcommand{\eeq}{\end{equation}}
\newcommand{\md}{{\rm d}}
\newcommand{\e}{{\rm e}}
\newcommand{\imu}{{\rm i}}

\newcommand{\bra}[1]{\langle #1 |}
\newcommand{\ket}[1]{| #1 \rangle}
\newcommand{\p}{{\scriptscriptstyle(+)}}
\newcommand{\m}{{\scriptscriptstyle(-)}}
\newcommand{\pa}{{\scriptscriptstyle\parallel}}
\newcommand{\pe}{{\scriptscriptstyle\perp}}
\newcommand{\T}{{\rm T}}
\newcommand{\msf}[1]{\mathsf{#1}}
\def\ga{\mathrel{\mathchoice {\vcenter{\offinterlineskip\halign{\hfil
$\displaystyle##$\hfil\cr>\cr\noalign{\vskip1.5pt}\sim\cr}}}
{\vcenter{\offinterlineskip\halign{\hfil$\textstyle##$\hfil\cr>\cr
\noalign{\vskip1.0pt}\sim\cr}}}
{\vcenter{\offinterlineskip\halign{\hfil$\scriptstyle##$\hfil\cr>\cr
\noalign{\vskip0.5pt}\sim\cr}}}
{\vcenter{\offinterlineskip\halign{\hfil$\scriptscriptstyle##$\hfil
\cr>\cr\noalign{\vskip0.5pt}\sim\cr}}}}}
\title{\vspace*{-2cm}\hspace*{\fill} {\normalsize SI-97-15}\\[3cm]
{\huge\bf Collinear Asymptotic Dynamics for Massive Particles.\\
Multi-Channel Regge Behaviour.}\thanks{This work has been supported 
under the German-Polish agreement on bilateral scientific and 
technological cooperation.}}
\author{ \\  \\ \\
\bf{J. Boguszynski, H. D. Dahmen, R. Kretschmer, L. {\L}ukaszuk}\\ \\ 
Siegen University, Fachbereich Physik, Siegen, Germany\\
and\\
Soltan Institute for Nuclear Studies, Warsaw, Poland\\ \\}
\date{October 1997}
\begin{document}
\maketitle
\vspace{5cm}

\begin{abstract}
\large{The high-energy behaviour in a multi-channel system is
investigated in the framework of collinear asymptotic dynamics for 
massive particles. We consider the most general trilinear coupling of 
$N$ different scalar fields. We find Regge behaviour and obtain closed 
expressions for the Regge trajectories and couplings. The results are 
corraborated in the multi-channel Bethe-Salpeter approach. The 
scattering processes dominated by single Regge trajectories are 
explicitly given.}
\end{abstract}

\newpage

\section{Introduction}
\label{s1}

The framework of collinear asymptotic dynamics for massive particles
\cite{bo} is a natural extension of methods that use large-time scale
Hamiltonians \cite{ku} to treat massless particles. The existence of
massless fields in the theory may generally demand a special treatment 
of infrared and collinear divergencies \cite{ku, ki}. In theories and
processes with massive particles collinear configurations play a role 
in the high-momentum regime only. In this context it has been shown 
that our approach \cite{bo} is suitable for the description of Regge
behaviour in the $\varphi^3$-theory. Since, to the best of
our knowledge, multi-channel Regge behaviour has not been studied
in the literature, we investigate the high-energy behaviour for the
most general trilinear coupling of $N$ hermitian scalar fields,
$\varphi_i$, $i = 1, \ldots, N$,
\beq\label{1a}
\mathcal{H}_\mathrm{I}(x) = - {1 \over 3!} \sum_{i, j, k = 1}^N 
c_{i j k} \, {:} \varphi_i(x) \varphi_j(x) \varphi_k(x) {:} \quad.
\eeq
The totally symmetric symbol $c_{i j k}$ denotes the coupling
constants.

Starting from a time-ordered exponential approximation of the
time-evolution operator \cite{bo}, we derive a matrix formulation for 
the multi-channel amplitudes at high energies. At large time scales 
the dynamics is dominated by collinear processes. In this regime
we obtain multi-channel Regge behaviour. 

We find a generally non-orthogonal base of states in the 
crossed channel leading to a separation of the various Regge 
trajectories. This fact implies a set of linear combinations of 
the amplitudes, the high-energy behaviour of which is expressed 
by one Regge trajectory only. 

The feasibility of this approach we find encouraging for further
studies of more realistic interactions. In this situation it is of
particular interest to confront our approach with a Bethe-Salpeter
formulation of the problem in the crossed channel. We generalize
methods successful for the single-channel system to the
multi-channel case.

The results of the Bethe-Salpeter treatment confirm the findings
of our $s$-channel treatment of the collinear asymptotic dynamics.

\section{Collinear Asymptotic Dynamics}
\label{s2}
\setcounter{equation}{0}

We consider the scattering matrix element
$\mathcal{T}_\mathrm{f{}i}((\vec{p}_1, r), (\vec{p}_2, a) \to 
(\vec{p}_3, s), (\vec{p}_4, b))$, where $r$, $a$, $s$ and $b$
denote the particle type. Let $p^{(j)} = (E^{(j)}(\vec{p}), \vec{p})$,
$E^{(j)}(\vec{p}) = \sqrt{\vec{p}^2 + m_j^2}$, stand for the 
four momentum and energy of a particle of type $j$, respectively. 
The scattering-matrix element can be written as
\begin{eqnarray}
\mathcal{T}_\mathrm{f{}i} & = & T + T^\mathrm{cr} \quad, \\
T & = & - \imu \int \md^4 x \, \md^4 y \, \e^{- \imu (p_2^{(a)} y -
p_4^{(b)} x)} \Theta(x^0 - y^0)
\bra{\vec{p}_3, s} j_{b \mathrm{H}}(x)
j_{a \mathrm{H}}(y) \ket{\vec{p}_1, r} \\
& = & - \imu \int \md^4 x \, \md^4 y \, \e^{- \imu (p_2^{(a)} y -
p_4^{(b)} x)} \Theta(x^0 - y^0) \label{2a} \\
& & \times \bra{\vec{p}_3, s} U^\dagger(x^0, - \infty) j_{b}(x)
U(x^0, y^0) j_{ a}(y) U(y^0, - \infty) \ket{\vec{p}_1, r} \quad, 
\nonumber
\end{eqnarray}
and
\[
T^\mathrm{cr} = T(a \leftrightarrow b, p_2^{(a)} \leftrightarrow -
p_4^{(b)}) \quad.
\]
In (\ref{2a}) the current for a particle of type $i$ is given by
\beq
j_i(x) = {1 \over 2!} \sum_{j, k} c_{i j k} \, {:} \varphi_j(x)
\varphi_k(x) {:} \quad.
\eeq

The matrix element $T$ will be calculated in the laboratory frame 
of the particle with momentum $p_2^{(a)}$; $p_1^{(r)}$ and 
$p_3^{(s)}$ are large momenta with small transfer 
$t = (p_1^{(r)} - p_3^{(s)})^2$.

As in our previous paper \cite{bo} we use as an asymptotic
approximation the Hamiltonian
\[
\tilde{H}_\mathrm{I} = H_\mathrm{I}^\p + H_\mathrm{I}^\m
\]
with
\beq
H_\mathrm{I}^\p = - {1 \over 2} \int \md^3 x \, \sum_{i, j, k} 
c_{i j k} \varphi_i^\m \varphi_j^\m \varphi_k^\p \quad,\quad
H_\mathrm{I}^\m = (H_\mathrm{I}^\p)^\dagger \quad.
\eeq
Here $\varphi_i^\p$ and $\varphi_i^\m$ are the annihilation and
creation parts of $\varphi_i$, respectively. As in the case of one 
particle type, we can replace $j_i(x)$ by
\beq\label{2k}
j_i^{(\mathrm{S})} = \sum_{j, k} c_{i j k} \varphi_j^\m
\varphi_k^\p \quad.
\eeq
Following the arguments given in \cite{bo}, we replace 
$U(x^0, y^0) = \mathbf{1}$ in (\ref{2a}) and insert a complete set of 
states between $j_b(x)$ and $j_a(y)$. In order to make the calculation
closely connected to the one in our previous paper, we use the 
following decomposition of the unit operator:
\begin{eqnarray}
\mathbf{1} & = & \ket{0} \bra{0} + \sum_{n = 1}^\infty {1 \over n!}
\sum_{i_1, \ldots, i_n = 1}^N \int \prod_{l = 1}^n 
{\md^3 k_l \over 2 (2 \pi)^3 E^{(i_l)}(\vec{k}_l)} \\
& & \times \ket{(\vec{k}_1, i_1), \ldots, (\vec{k}_n, i_n)}
\bra{(\vec{k}_1, i_1), \ldots, (\vec{k}_n, i_n)} \quad. \nonumber
\end{eqnarray}
The validity of this decomposition can be readily verified. 
Using this expression, we get
\begin{eqnarray}
T & = & (2 \pi)^4 \delta(p_1^{(r)} + p_2^{(a)} - p_3^{(s)} -
p_4^{(b)}) \mathcal{M}_{s, b}^{r, a} \quad, \\
\mathcal{M}_{s, b}^{r, a} & = & (2 \pi)^3 \sum_{n = 1}^\infty 
{1 \over n!} \sum_{i_1, \ldots, i_n = 1}^N \int \prod_{l = 1}^n 
{\md^3 k_l \over 2 (2\pi)^3 E^{(i_l)}(\vec{k}_l)} \\
& & \times {\delta(\vec{p}_1 + \vec{p}_2 
- {\sum_{l = 1}^n \vec{k}_l}) \over E^{(r)}(\vec{p}_1) 
+ E^{(a)}(\vec{p}_2) - {\sum_{l = 1}^n E^{(i_l)}(\vec{k}_l)} 
+ \imu \varepsilon} \nonumber \\
& & \times G_a^n((\vec{p}_1, r); (\vec{k}_1, i_1), \ldots,
(\vec{k}_n, i_n))
G_b^{n \ast}((\vec{p}_3, s); (\vec{k}_1, i_1), \ldots,
(\vec{k}_n, i_n)) \nonumber
\end{eqnarray}
with
\begin{eqnarray}
\lefteqn{G_a^n((\vec{p}, i); 
(\vec{k}_1, i_1), \ldots, (\vec{k}_n, i_n))
= \bra{(\vec{k}_1, i_1), \ldots, (\vec{k}_n, i_n)}
j_a^{(\mathrm{S})}(0) U(0, - \infty) \ket{\vec{p}, i}} 
\nonumber \\
& = & \sum_{\nu = 1}^n \sum_d c_{a i_\nu d}
\bra{(\vec{k}_1, i_1), \ldots, \widehat{(\vec{k}_\nu, i_\nu)}, \ldots,
(\vec{k}_n, i_n)} \varphi_d^\p(0) U(0, - \infty)
\ket{\vec{p}, i} \quad,
\end{eqnarray}
where the symbol $\widehat{(\vec{k}_\nu, i_\nu)}$ indicates that the 
corresponding particle has to be omitted from the state.

As in our previous paper, the time-evolution operator $U(0, - \infty)$ 
is dominated by the operator
\beq
W_1(0, - \infty) = \T \exp \left\{ - \imu \int\limits_{- \infty}^0 
H_\mathrm{I}^\p(t) \, \md t \right\} \quad.
\eeq
In the time-ordered expansion of $W_1$ only the term of order $n -1$ 
contributes to $G_a^n$, i.~e.
\begin{eqnarray}
\lefteqn{\bra{(\vec{k}_1, i_1), \ldots, 
\widehat{(\vec{k}_\nu, i_\nu)}, \ldots, (\vec{k}_n, i_n)} 
\varphi_d^\p(0) W_1(0, - \infty)
\ket{\vec{p}, i}} \label{2b} \\
& = & (- \imu)^{n - 1} \int\limits_{- \infty}^0 \md t_1 \, \cdots \, 
\md t_{n - 1} \, \Theta(t_1, \ldots, t_{n - 1}) \nonumber \\
& & \times \bra{0} \varphi_d^\p(0) a_{i_1}(\vec{k}_1) \cdots
\widehat{a_{i_\nu}(\vec{k}_\nu)} \cdots a_{i_n}(\vec{k}_n) 
H_\mathrm{I}^\p(t_1) \cdots H_\mathrm{I}^\p(t_{n - 1}) 
\ket{\vec{p}, i} \quad. \nonumber
\end{eqnarray}
In this expression, $\Theta(t_1, \ldots, t_{n - 1}) =
\Theta(t_1 - t_2) \cdots \Theta(t_{n - 2} - t_{n - 1})$, and
$a_i(\vec{k})$ is the annihilation operator of a particle of type $i$. 
With the definition
\beq
D_i(t, \vec{k}) = - \imu [a_i(\vec{k}), H_\mathrm{I}^\p(t)] 
= \imu \int_{x^0 = t} \md^3 x \, \e^{\imu k^{(i)} x} 
j_i^{(\mathrm{S})}(x) \quad,
\eeq
cf.\ (\ref{2k}), the dominant contribution to (\ref{2b}) can be written 
as
\begin{eqnarray}
& & \int\limits_{- \infty}^0 \md t_1 \, \cdots \, 
\md t_{n - 1} \, \Theta(t_1, \ldots, t_{n - 1}) \nonumber \\
& & \times \sum_{\pi_\nu} \bra{0} \varphi_d^\p(0) 
D_{i_{\pi(1)}}(t_1, \vec{k}_{\pi(1)})
D_{i_{\pi(2)}}(t_2, \vec{k}_{\pi(2)}) \cdots
D_{i_{\pi(n)}}(t_{n - 1}, \vec{k}_{\pi(n)}) \ket{\vec{p}, i} 
\quad, \nonumber
\end{eqnarray}
where the sum runs over all permutations of the integers 1, 2, 
\ldots, $\nu - 1$, $\nu + 1$, \ldots, $n$. Since
\beq
D_i(t, \vec{k}) \ket{\vec{p}, j} = \imu \sum_{l} c_{i j l}
{\e^{\imu (E^{(l)}(\vec{p} - \vec{k}) + E^{(i)}(\vec{k}) - 
E^{(j)}(\vec{p})) t} \over 
2 E^{(l)}(\vec{p} - \vec{k})} \ket{\vec{p} - \vec{k}, l} \quad,
\eeq
the result for $G_a^n$ is
\begin{eqnarray}
\lefteqn{G_a^n((\vec{p}, r); (\vec{k}_1, i_1), \ldots,
(\vec{k}_n, i_n))} \label{2c} \\
& = & \sum_{\nu = 1}^n \sum_{\pi_\nu} 
\sum_{u_1, \ldots, u_{\nu - 1}, \atop u_{\nu + 1}, \ldots, u_n}
{c_{r i_{\pi(n)} u_n} \over 
(\vec{p}^\pe - \vec{q}_{n, \nu}^\pe)^2 + m_{u_n}^2}
{c_{u_n i_{\pi(n - 1)} u_{n - 1}} \over
(\vec{p}^\pe - \vec{q}_{n - 1, \nu}^\pe)^2 + m_{u_{n - 1}}^2} 
\cdots \nonumber \\
& & \times 
{c_{u_{\nu + 2} i_{\pi(\nu + 1)} u_{\nu + 1}} \over
(\vec{p}^\pe - \vec{q}_{\nu + 1, \nu}^\pe)^2 + m_{u_{\nu + 1}}^2}
{c_{u_{\nu + 1} i_{\pi(\nu - 1)} u_{\nu - 1}} \over
(\vec{p}^\pe - \vec{q}_{\nu - 1, \nu}^\pe)^2 + m_{u_{\nu - 1}}^2}
\cdots
{c_{u_2 i_{\pi(1)} u_1} \over
(\vec{p}^\pe - \vec{q}_{1, \nu}^\pe)^2 + m_{u_1}^2}
c_{u_1 i_\nu a} \nonumber
\end{eqnarray}
with
\[
\vec{q}_{i, \nu}^\pe = \sum_{j = i, \, j \neq \nu}^n
\vec{k}_{\pi(j)}^\pe \quad,
\]
and
\beq\label{2d}
p^\pa \ga k_{\pi(n)}^\pa \gg k_{\pi(n - 1)}^\pa \gg \ldots \gg 
k_{\pi(\nu + 1)}^\pa \gg k_{\pi(\nu - 1)}^\pa \gg \ldots \gg
k_{\pi(1)}^\pa \gg k_\nu^\pa \gg o(m) \quad.
\eeq
Equations (\ref{2c}) and (\ref{2d}) are a consequence 
of collinear asymptotic dynamics \cite{bo}.

In what follows, the decomposition of momenta in longitudinal and 
transverse parts is to be done with respect to $\vec{p}_1$.
In this way we obtain
\begin{eqnarray}
\lefteqn{\mathcal{M}} \label{2e} \\
& = & (2 \pi)^3 \sum_{n = 1}^\infty {1 \over n!}
\sum_{i_1, \ldots, i_n = 1}^N \int \prod_{l = 1}^n
{\md k_l^\pa \, \md^2 k_l^\pe
\over 2 (2 \pi)^3 E^{(i_l)}(\vec{k}_l^\pa)}
{\delta(\vec{p}_1 - \sum_{l = 1}^n \vec{k}_l)
\over E^{(r)}(p_1^\pa) + m_a - \sum_{l = 1}^n E^{(i_l)}(k_l^\pa)} 
\nonumber \\
& & \times \Biggl( \sum_{\nu = 1}^n \sum_{\pi \in \pi_\nu}
\Theta(k_{\pi(n)}^\pa, \ldots, k_{\pi(1)}^\pa, k_\nu^\pa)
\sum_{u_1, \ldots, u_{\nu - 1}, \atop u_{\nu + 1}, \ldots, u_n}
{c_{r i_{\pi(n)} u_n} \over
\vec{q}_{n, \nu}^{\pe 2} + m_{u_n}^2}
{c_{u_n i_{\pi(n - 1)} u_{n - 1}} \over
\vec{q}_{n - 1, \nu}^{\pe 2} + m_{u_{n - 1}}^2}
\cdots \nonumber \\
& & \times 
{c_{u_{\nu + 2} i_{\pi(\nu + 1)} u_{\nu + 1}} \over
\vec{q}_{\nu + 1, \nu}^{\pe 2} + m_{u_{\nu + 1}}^2}
{c_{u_{\nu + 1} i_{\pi(\nu - 1)} u_{\nu - 1}} \over
\vec{q}_{\nu - 1, \nu}^{\pe 2} + m_{u_{\nu - 1}}^2}
\cdots
{c_{u_2 i_{\pi(1)} u_1} \over
\vec{q}_{1, \nu}^{\pe 2} + m_{u_1}^2}
c_{u_1 i_\nu a} 
\Biggr) \nonumber \\
& & \times \Biggl( \sum_{\mu = 1}^n \sum_{\pi' \in \pi_\mu}
\Theta(k_{\pi'(n)}^\pa, \ldots, k_{\pi'(1)}^\pa, k_\mu^\pa)
\sum_{w_1, \ldots, w_{\mu - 1}, \atop w_{\mu + 1}, \ldots, w_n}
{c_{s i_{\pi'(n)} w_n} \over
(\vec{p}_3^\pe - \vec{q}_{n, \mu}^\pe)^2 + m_{w_n}^2}
\nonumber \\
& & \times {c_{w_n i_{\pi'(n - 1)} w_{n - 1}} \over
(\vec{p}_3^\pe - \vec{q}_{n - 1, \mu}^\pe)^2 + m_{w_{n - 1}}^2}
\cdots 
{c_{w_2 i_{\pi'(1)} w_1} \over
(\vec{p}_3^\pe - \vec{q}_{1, \mu}^\pe)^2 + m_{w_1}^2}
c_{w_1 i_\mu b} 
\Biggr) \quad. \nonumber
\end{eqnarray}
Note that this expression factorizes into longitudinal and 
transverse parts. Because of the strong ordering (\ref{2d}) 
only the same permutations of momenta in both factors of 
(\ref{2e}) give a contribution to $\mathcal{M}$. Therefore we get
\begin{eqnarray*}
\lefteqn{\mathcal{M}} \\
& = & (2 \pi)^3 \sum_{n = 1}^\infty {1 \over n!}
\sum_{i_1, \ldots, i_n = 1}^N \int \prod_{l = 1}^n
{\md k_l^\pa \, \md^2 k_l^\pe
\over 2 (2 \pi)^3 E^{(i_l)}(\vec{k}_l^\pa)}
{\delta(\vec{p}_1 - \sum_{l = 1}^n \vec{k}_l)
\over E^{(r)}(p_1^\pa) + m_a - \sum_{l = 1}^n E^{(i_l)}(k_l^\pa)} \\
& & \times \sum_{\nu = 1}^n \sum_{\pi_\nu}
\Theta(k_{\pi(n)}^\pa, \ldots, k_{\pi(1)}^\pa, k_\nu^\pa) \!\!
\sum_{u_1, \ldots, u_{\nu - 1}, \atop u_{\nu + 1}, \ldots, u_n}
\sum_{w_1, \ldots, w_{\nu - 1}, \atop w_{\nu + 1}, \ldots, w_n} 
{c_{r i_{\pi(n)} u_n} c_{s i_{\pi(n)} w_n} \over
(\vec{q}_{n, \nu}^{\pe 2} + m_{u_n}^2)
((\vec{p}_3^\pe - \vec{q}_{n, \nu}^\pe)^2 + m_{w_n}^2)} \\
& & \times \cdots {c_{u_2 i_{\pi(1)} u_1} c_{w_2 i_{\pi(1)} w_1} \over
(\vec{q}_{1, \nu}^{\pe 2} + m_{u_1}^2)
((\vec{p}_3^\pe - \vec{q}_{1, \nu}^\pe)^2 + m_{w_1}^2)} 
c_{u_1 i_\nu a} c_{w_1 i_\nu b} \quad.
\end{eqnarray*}
In this form, the integration over transverse momenta can be
factorized:
\begin{eqnarray*}
\mathcal{M} & = & {1 \over 2} \sum_{n = 1}^\infty {1 \over n!}
\sum_{i_1, \ldots, i_n = 1}^N \int \prod_{l = 1}^n
{\md k_l^\pa \over E^{(i_l)}(\vec{k}_l^\pa)}
{\delta(\vec{p}_1^\pa - \sum_{l = 1}^n \vec{k}_l^\pa)
\over E^{(r)}(p_1^\pa) + m_a - \sum_{l = 1}^n E^{(i_l)}(k_l^\pa)} \\
& & \times \sum_{\nu = 1}^n \sum_{\pi_\nu}
\Theta(k_{\pi(n)}^\pa, \ldots, k_{\pi(1)}^\pa, k_\nu^\pa)
\sum_{u_1, \ldots, u_{\nu - 1}, \atop u_{\nu + 1}, \ldots, u_n}
\sum_{w_1, \ldots, w_{\nu - 1}, \atop w_{\nu + 1}, \ldots, w_n} 
c_{r i_{\pi(n)} u_n} c_{s i_{\pi(n)} w_n} \\
& & \times F^{u_n}_{w_n}(\vec{p}_3^\pe) \cdots
c_{u_2 i_{\pi(1)} u_1} c_{w_2 i_{\pi(1)} w_1}
F^{u_1}_{w_1}(\vec{p}_3^\pe)
c_{u_1 i_\nu a} c_{w_1 i_\nu b} \quad,
\end{eqnarray*}
where
\beq\label{2l}
F^u_w(\vec{p}_3^\pe) = {1 \over 2 (2 \pi)^3} \int
{\md^2 q^\pe \over (\vec{q}^{\pe 2} + m_u^2) 
((\vec{p}_3^\pe - \vec{q}^\pe)^2 + m_w^2)} \quad.
\eeq
The longitudinal integral can be carried out in leading logarithmic 
approximation, resulting in ($s = 2 m_a p_1^\pa$)
\beq
\int\limits_0^\infty \prod_{l = 1}^n 
{\md k_l^\pa \over E^{(i_l)}(k_l^\pa)} 
{\delta(p_1^\pa - \sum_{l = 1}^n k_l^\pa)
\Theta(k_{\pi(n)}^\pa, \ldots, k_{\pi(1)}^\pa, k_\nu^\pa)
\over E^{(r)}(p_1^\pa) + m_a - \sum_{l = 1}^n E^{(i_l)}(k_l^\pa)}
= {2 \over s} {(\ln s)^{n - 1} \over (n - 1)!}
\eeq
for all permutations. Thus all permutations for all $\nu$ give the same
contribution to $\mathcal{M}$, so that, if we define
\beq\label{2h}
C_{w_1, w_2}^{u_1, u_2} = \sum_{i = 1}^N c_{u_1 i u_2} c_{w_1 i w_2} 
\quad,
\eeq
the matrix element can be written as
\beq
\mathcal{M} = \mathcal{M}^{r, a}_{s, b}
= {1 \over s} \sum_{n = 1}^\infty {(\ln s)^{n - 1} \over (n - 1)!} 
\sum_{u_1, \ldots, u_{n -1}, \atop w_1, \ldots, w_{n - 1}}
C_{s, w_{n - 1}}^{r, u_{n -1}} F_{w_{n - 1}}^{u_{n - 1}} 
C_{w_{n - 1}, w_{n - 2}}^{u_{n - 1}, u_{n - 2}}
\cdots C_{w_2, w_1}^{u_2, u_1} F_{w_1}^{u_1} 
C_{w_1, b}^{u_1, a} \quad.
\eeq
This formula has a simple interpretation in terms of processes in the
crossed channel. To see this, we introduce an index $I$,
\[
(r, s) \to I(r,s) = 1, \ldots, N^2 \quad,
\]
characterizing the states in the crossed channel, and define matrices
\begin{eqnarray}
\msf{M}_{I(r, s) I(a, b)} & = & \mathcal{M}_{s, b}^{r, a} \quad, \\
\msf{C}_{I(r, s) I(a, b)} & = & C_{s, b}^{r, a} \quad, \label{2g} \\
\msf{F}_{I J} & = & \delta_{I J} F_I \quad,\quad 
F_{I(u, w)} = F_w^u \quad. \label{2i}
\end{eqnarray}
With this, we get
\beq\label{2f}
\msf{M} = {1 \over s} \msf{C} \exp (\msf{F} \msf{C} \ln s)
= {1 \over s} \exp (\msf{C} \msf{F} \ln s) \msf{C}
= {1 \over s} \sqrt{\msf{F}}^{- 1} \left( \tilde{\msf{C}} 
\exp (\tilde{\msf{C}} \ln s) \right) \sqrt{\msf{F}}^{- 1}
\quad, 
\eeq
$\tilde{\msf{C}} = \sqrt{\msf{F}} \msf{C} \sqrt{\msf{F}}$.

\subsection*{Example}

Consider the interaction
\beq\label{2j}
\mathcal{H}_\mathrm{I} = - {g_1 \over 2!} \, {:} \varphi^2 \sigma {:}
- {g_2 \over 3!} \, {:} \sigma^3 {:} \quad,
\eeq
which is a special case of (\ref{1a}) for $N = 2$, 
$\varphi_1 = \varphi$, $\varphi_2 = \sigma$, and
$c_{112} = c_{121} = c_{211} = g_1$, $c_{222} = g_2$,
$c_{111} = c_{122} = c_{212} = c_{221} = 0$. Let
$I(1, 1) = 1$, $I(2, 2) = 2$, $I(1, 2) = 3$, and $I(2, 1) = 4$, then 
the matrix $\msf{C}$ is block diagonal,
\[
\msf{C} = \left( \begin{array}{cc}
	\msf{C}' & 0 \\
	0 & \msf{C}''
\end{array} \right) \quad,\quad
\msf{C}' = \left( \begin{array}{cc}
	 g_1^2 & g_1^2 \\
	 g_1^2 & g_2^2 
\end{array} \right) \quad,\quad
\msf{C}'' = \left( \begin{array}{cc}
	 g_1 g_2 & g_1^2 \\
	 g_1^2 & g_1 g_2 
\end{array} \right) \quad,
\]
and $\msf{F}$ is diagonal with $F_1 = F_1^1$, $F_2 = F_2^2$, 
$F_3 = F_4 = F_1^2 = F_2^1$, where $F^u_w$ is given by (\ref{2l}).
The symmetric matrix $\tilde{\msf{C}}$ is 
\[
\tilde{\msf{C}} = \left( \begin{array}{cc}
	\tilde{\msf{C}}' & 0 \\
	0 & \tilde{\msf{C}}''
\end{array} \right) \quad,\quad
\tilde{\msf{C}}' = \left( \begin{array}{cc}
	F_1 g_1^2 & \sqrt{F_1 F_2} g_1^2 \\
	\sqrt{F_1 F_2} g_1^2 & F_2 g_2^2
\end{array} \right) \quad,\quad
\tilde{\msf{C}}'' = F_3 \msf{C}'' \quad.
\]
Therefore, also $\msf{M}$ is block diagonal. Let us only consider the
upper left part, which is given by
\[
\msf{M}' = {1 \over s} \sqrt{\msf{F}'}^{- 1} \tilde{\msf{C}}'
\exp(\tilde{\msf{C}}' \ln s) \sqrt{\msf{F}'}^{- 1} \quad.
\]
By diagonalizing $\tilde{\msf{C}}'$, it is easy to see that,
e.~g.,
\begin{eqnarray}
\lefteqn{\mathcal{M}_{\varphi, \varphi \to \varphi, \varphi} 
= \msf{M}_{1 1}'} \nonumber \\
& = & {g_1^2 \over \lambda^{[1]} - \lambda^{[2]}} 
\left[ \left( \lambda^{[1]} + F_2 (g_1^2 - g_2^2) \right) 
s^{\lambda^{[1]} - 1}
- \left( \lambda^{[2]} + F_2 (g_1^2 - g_2^2) \right) 
s^{\lambda^{[2]} - 1} \right] \, ,
\end{eqnarray}
with the eigenvalues
\beq\label{2m}
\lambda^{[1, 2]} = {1 \over 2} \left( F_1 g_1^2 + F_2 g_2^2
\pm \sqrt{(F_1 g_1^2 - F_2 g_2^2)^2 + 4 F_1 F_2 g_1^4}
\right)
\eeq
of $\tilde{\msf{C}}'$. Thus the high-energy behaviour of 
$\mathcal{M}_{\varphi, \varphi \to \varphi, \varphi}$ is governed by 
two Regge terms.

We will come back to this example in sec.~\ref{s4}.
 
\section{Multi-Channel Bethe-Salpeter Equation}
\label{s3}
\setcounter{equation}{0}

Equation (\ref{2f}) shows that the amplitude $T$ possesses
an exponential $\ln s$ dependence, which is a typical feature of 
Regge behaviour. In order to see that this result really reflects 
the existence of Regge trajectories in the crossed channel, we will 
now formulate the multi-channel Bethe-Salpeter equation and study its 
solution. Although the one-channel Bethe-Salpeter equation was 
considered by Lee and Sawyer \cite{ls} and many others \cite{na} long 
ago, to the best of our knowledge there is no treatment of the 
multi-channel case. 

Let us start with the off-shell amplitude
\beq
T((q_1, i_1), (q_3, i_3)  \to (q_2, i_2), (q_4, i_4)) = (2 \pi)^4 
\delta(q_1 + q_3 - q_2 - q_4) 
t_{i_3, i_4}^{i_1, i_2}(P_\mathrm{tot}, q_{1 3}, q_{2 4})
\eeq
with $P_\mathrm{tot} = q_1 + q_3 = q_2 + q_4$ and
$q_{i j} = (q_i - q_j) / 2$.

The Bethe-Salpeter equation for 
$t_{i_3, i_4}^{i_1, i_2}(P_\mathrm{tot}, q_{1 3}, q_{2 4})$ with the
interaction (\ref{1a}) in the ladder approximation reads
\begin{eqnarray}
\lefteqn{t_{i_3, i_4}^{i_1, i_2}(P_\mathrm{tot}, q_{1 3}, q_{2 4})}
\label{3a} \\ 
& = & \sum_i {c_{i_1 i i_2} c_{i_3 i i_4} \over
(q_{1 3} - q_{2 4})^2 - m_i^2 + \imu \varepsilon}
+ {\imu \over (2 \pi)^4} \sum_{i, j, k} \int \md^4 q \,
{c_{i_1 i j} c_{i_3 i k} \over
(q_{1 3} - q)^2 - m_i^2 + \imu \varepsilon} \nonumber \\
& & \times {1 \over (P_\mathrm{tot} / 2 + q)^2 - m_j^2 
+ \imu \varepsilon}
{1 \over (P_\mathrm{tot} / 2 - q)^2 - m_k^2 + \imu \varepsilon}
t_{k, i_4}^{j, i_2}(P_\mathrm{tot}, q, q_{2 4}) \quad. \nonumber
\end{eqnarray}
In the c.~m.\ frame ($P_\mathrm{tot}^\mathrm{cm} = (W, 0)$) the 
partial-wave decomposition of $t_{i_3, i_4}^{i_1, i_2}$ is given by
\beq\label{3b}
t_{i_3, i_4}^{i_1, i_2}(P_\mathrm{tot}^\mathrm{cm}, q_{1 3}, q_{2 4})
= {1 \over 4 \pi |\vec{q}_{1 3}| |\vec{q}_{2 4}|} 
\sum_{\ell = 0}^\infty (2 \ell + 1) P_\ell(\cos \vartheta)
t_\ell^{i_1, i_3; i_2, i_4}(W, q_{1 3}^0, |\vec{q}_{1 3}|, 
q_{2 4}^0, |\vec{q}_{2 4}|) 
\eeq
($\vartheta$ is the angle between $\vec{q}_{1 3}$ and $\vec{q}_{2 4}$), 
so that
\beq\label{3c}
t_\ell^{i_1, i_3; i_2, i_4}(W, q_{1 3}^0, |\vec{q}_{1 3}|, 
q_{2 4}^0, |\vec{q}_{2 4}|)
= 2 \pi |\vec{q}_{1 3}| |\vec{q}_{2 4}| \int \md \cos \vartheta \, 
P_\ell(\cos \vartheta) 
t_{i_3, i_4}^{i_1, i_2}(P_\mathrm{tot}^\mathrm{cm}, q_{1 3}, q_{2 4}) 
\quad. \nonumber
\eeq
Then using (\ref{3c}) and (\ref{3a}), we arrive at the Bethe-Salpeter
equation for the partial-wave amplitude,
\begin{eqnarray}
\lefteqn{t_\ell^{i_1, i_3; i_2, i_4}(W, q_{1 3}^0, |\vec{q}_{1 3}|, 
q_{2 4}^0, |\vec{q}_{2 4}|)} \label{3d} \\
& = & - 2 \pi \sum_i c_{i_1 i i_2} c_{i_3 i i_4} 
Q_\ell(\beta_i(q_{1 3}^0, |\vec{q}_{1 3}|, q_{2 4}^0, |\vec{q}_{2 4}|)) 
- {\imu \over (2 \pi)^3} \sum_{i, j, k} \int \md q^0 \, \md |\vec{q}| 
\nonumber \\
& & \times {Q_\ell(\beta_i(q_{1 3}^0, |\vec{q}_{1 3}|, 
q^0, |\vec{q}|))
c_{i_1 i j} c_{i_3 i k} t_\ell^{j, k; i_2, i_4}(W, q^0, |\vec{q}|, 
q_{2 4}^0, |\vec{q}_{2 4}|) \over
[(P_\mathrm{tot}^\mathrm{cm} / 2 + q)^2 - m_j^2 + \imu \varepsilon]
[(P_\mathrm{tot}^\mathrm{cm} / 2 - q)^2 - m_k^2 + \imu \varepsilon]}
\quad, \nonumber
\end{eqnarray}
where $Q_\ell$ is a Legendre function of the second kind, and its
argument is given by
\[
\beta_i(q_{1 3}^0, |\vec{q}_{1 3}|, q_{2 4}^0, |\vec{q}_{2 4}|) 
= {- (q_{1 3}^0 - q_{2 4}^0)^2 + \vec{q}_{1 3}^2 + \vec{q}_{2 4}^2 
+ m_i^2 - \imu \varepsilon \over
2 |\vec{q}_{1 3}| |\vec{q}_{2 4}|} \quad.
\]

As we are only interested in the Regge trajectories in the leading 
logarithmic approximation and near $\ell = - 1$, we adopt the method 
used in \cite{ls} for the one-channel case, which amounts to replacing 
$Q_\ell$ by its leading $\ell$-plane singularity,
\[
Q_\ell(\beta) \to {1 \over \ell + 1} \quad.
\]
Inserting this into (\ref{3d}) and using the matrix notation for
products of $c_{i j k}$ introduced in (\ref{2g}) and (\ref{2h}),
\[
\msf{C}_{I(i, j) I(k, l)} = \sum_{m = 1}^N c_{i m k} c_{j m l} \quad,
\]
and
\beq
{\msf{t}_\ell}_{I(i, j) I(k, l)} = t_\ell^{i, j; k, l} \quad,
\eeq
we end up with a separable form of the Bethe-Salpeter equation,
\begin{eqnarray}
\lefteqn{(\ell + 1) {\msf{t}_\ell}_{I J}(W, q_{1 3}^0, |\vec{q}_{1 3}|, 
q_{2 4}^0, |\vec{q}_{2 4}|)} \\ 
& = & - 2 \pi \msf{C}_{I J} - {\imu \over (2 \pi)^3} \sum_{K = 1}^{N^2} 
\int \md q^0 \, \md |\vec{q}| \, \msf{C}_{I K} f_K(W, q^0, |\vec{q}|) 
{\msf{t}_\ell}_{K J}(W, q^0, |\vec{q}|, q_{2 4}^0, |\vec{q}_{2 4}|) 
\quad, \nonumber
\end{eqnarray}
where
\[
f_{I(j, k)}(W, q^0, |\vec{q}|) = {1 \over 
(P_\mathrm{tot}^\mathrm{cm} / 2 + q)^2 - m_j^2 + \imu \varepsilon} 
{1 \over (P_\mathrm{tot}^\mathrm{cm} / 2 - q)^2 - m_k^2 
+ \imu \varepsilon} \quad.
\]
In matrix form, the solution of this equation is
\beq
\msf{t}_\ell = - 2 \pi \msf{C} {1 \over \ell + 1 - \hat{\msf{F}} 
\msf{C}} 
= - 2 \pi {\textstyle \sqrt{\hat{\msf{F}}}}^{- 1} \hat{\msf{C}}
{1 \over \ell + 1 - \hat{\msf{C}}}
{\textstyle \sqrt{\hat{\msf{F}}}}^{- 1} \quad,
\eeq
$\hat{\msf{C}} = \sqrt{\hat{\msf{F}}} \msf{C} \sqrt{\hat{\msf{F}}}$. 
Here the matrix $\hat{\msf{F}}$ is given by
\beq
\hat{\msf{F}}_{I J}(W) = {- \imu \delta_{I J} \over (2 \pi)^3}
\int \md q^0 \, \md |\vec{q}| \,
f_I(W, q^0, |\vec{q}|) \quad.
\eeq

Since $\hat{\msf{C}}$ is symmetric, we can diagonalize it: Let
$\msf{v}^{[i]}$ denote the normalized eigenvectors,
\beq
\hat{\msf{C}} \msf{v}^{[i]} = \lambda^{[i]} \msf{v}^{[i]} \quad,\quad
i = 1, \ldots, N^2 \quad,
\eeq
then
\beq\label{3f}
\msf{t}_\ell = - 2 \pi {\textstyle \sqrt{\hat{\msf{F}}}}^{- 1} 
\left( \sum_i {\msf{v}^{[i]} \lambda^{[i]} \msf{v}^{[i] \dagger} \over
\ell + 1 - \lambda^{[i]}} \right) 
{\textstyle \sqrt{\hat{\msf{F}}}}^{- 1} 
\quad.
\eeq
Thus, the Regge trajectories are given by
\beq
\alpha^{[i]} = \ell_\mathrm{pole} = \lambda^{[i]} - 1 \quad.
\eeq
With (\ref{3f}) we are ready to use the 
Mandelstam-Sommerfeld-Watson representation \cite{ls, co, ma}
to obtain the high-energy amplitude in the crossed channel
($t = (q_1 + q_3)^2 = W^2$ fixed, $s = (q_1 - q_2)^2 \to \infty$),
with the result
\beq\label{3e}
\msf{t}_\mathrm{poles} 
\stackrel{{\scriptstyle s \to \infty,} 
\atop {\scriptstyle t \, \mathrm{fixed}}}{=} 
{1 \over s} {\textstyle \sqrt{\hat{\msf{F}}}}^{- 1} 
\left( \sum_i \msf{v}^{[i]} \lambda^{[i]} s^{\lambda^{[i]}} 
\msf{v}^{[i] \dagger} \right) 
{\textstyle \sqrt{\hat{\msf{F}}}}^{- 1}
= {1 \over s} {\textstyle \sqrt{\hat{\msf{F}}}}^{- 1} 
\left( \hat{\msf{C}} \exp (\hat{\msf{C}} \ln s) \right) 
{\textstyle \sqrt{\hat{\msf{F}}}}^{- 1} 
\quad. 
\eeq
It can be shown by elementary integration that $\hat{\msf{F}}$ in
(\ref{3e}) and $\msf{F}$ in (\ref{2f}) coincide. This shows that the
approaches of sec.~\ref{s2} and the present section are equivalent.

\section{Base States}
\label{s4}
\setcounter{equation}{0}

The states $\msf{v}^{[i]}$ are eigenstates of the hermitian matrix
$\hat{\msf{C}} = \tilde{\msf{C}}$, thus they are orthogonal. This is, 
however, no longer true for the states 
$\sqrt{\msf{F}}^{- 1} \msf{v}^{[i]}$. As the dual base for these 
states we introduce the states
\beq
\msf{u}^{[i]} = \sqrt{\msf{F}} \msf{v}^{[i]} \quad \mbox{with} \quad
\msf{u}^{[i] \dagger} \msf{F}^{- 1} \msf{u}^{[j]} = \delta_{i j} \quad.
\eeq
The only matrix elements of $\msf{M} = \msf{t}_\mathrm{poles}$ in 
this base that are non-zero are
\[
\msf{u}^{[i] \dagger} \msf{M} \msf{u}^{[i]}
\quad,\quad i = 1, \ldots, N^2 \quad.
\]
The high-energy behaviour of each of these matrix elements is 
governed by only one trajectory,
\beq\label{4a}
\msf{u}^{[i] \dagger} \msf{M} \msf{u}^{[j]} 
= \delta_{i j} \lambda^{[i]} s^{\lambda^{[i]} - 1}  \quad,
\eeq
although the states $\msf{u}^{[i]}$ are in general not orthogonal.
Equation (\ref{4a}) leads to the following sum rules for the scattering
amplitudes:
\beq\label{4b}
\sum_{I, J = 1}^{N^2} \msf{v}_I^{[i]} \msf{v}_J^{[j]} \sqrt{F_I F_J}
\msf{M}_{I J} = \delta_{i j} \lambda^{[i]} s^{\lambda^{[i]} - 1}
\quad.
\eeq

Note that some of these relations simplify whenever 
$\lambda^{[i]} = 0$. In this case,
\beq
\msf{M} \msf{u}^{[i]} = 0 \quad \mbox{for } \lambda^{[i]} = 0
\eeq
(or equivalently $\msf{C} \msf{u}^{[i]} = 0$), i.~e.\ states 
corresponding to fixed Regge singularities are transparent in our
approximation.

As an illustration, we again consider the example of sec.~\ref{s2},
eq.~(\ref{2j}):

The four eigenvalues are given by (\ref{2m}) and
\beq
\lambda^{[3,4]} = F_3 g_1 (g_2 \pm g_1) \quad,
\eeq
the corresponding eigenvectors are
\[
\msf{v}^{[1]} = {1 \over \sqrt{d}} \left( \begin{array}{c}
	\sqrt{a} \\ \sqrt{b} \\ 0 \\ 0
\end{array} \right) \, ,\,
\msf{v}^{[2]} = {1 \over \sqrt{d}} \left( \begin{array}{r}
	\sqrt{b} \\ - \sqrt{a} \\ 0 \\ 0
\end{array} \right) \, ,\,
\msf{v}^{[3]} = {1 \over \sqrt{2}} \left( \begin{array}{c}
	0 \\ 0 \\ 1 \\ 1
\end{array} \right) \, ,\,
\msf{v}^{[4]} = {1 \over \sqrt{2}} \left( \begin{array}{r}
	0 \\ 0 \\ 1 \\ - 1
\end{array} \right) \, ,
\]
with $d = \sqrt{(F_1 g_1^2 - F_2 g_2^2)^2 + 4 F_1 F_2 g_1^4}$,
$a = (d + F_1 g_1^2 - F_2 g_2^2) / 2$, and
$b = (d - F_1 g_1^2 + F_2 g_2^2) / 2$. Out of the ten relations 
described by (\ref{4b}), four are trivial due to the fact that 
$\msf{M}$ is block diagonal. The remaining relations read (using 
the symmetry of $\msf{M}$)
\beq\label{4c}
a F_1 \mathcal{M}_{\varphi, \varphi \to \varphi, \varphi}
+ 2 F_1 F_2 g_1^2
\mathcal{M}_{\varphi, \sigma \to \varphi, \sigma}
+ b F_2 \mathcal{M}_{\sigma, \sigma \to \sigma, \sigma}
= (\lambda^{[1]} - \lambda^{[2]}) \lambda^{[1]} 
s^{\lambda^{[1]} - 1} \quad, 
\eeq
\beq
b F_1 \mathcal{M}_{\varphi, \varphi \to \varphi, \varphi}
- 2 F_1 F_2 g_1^2
\mathcal{M}_{\varphi, \sigma \to \varphi, \sigma}
+ a F_2 \mathcal{M}_{\sigma, \sigma \to \sigma, \sigma}
= (\lambda^{[1]} - \lambda^{[2]}) \lambda^{[2]} 
s^{\lambda^{[2]} - 1} \quad,
\eeq
\beq
\mathcal{M}_{\varphi, \varphi \to \sigma, \sigma}
+ 2 \mathcal{M}_{\varphi, \sigma \to \sigma, \varphi}
+ \mathcal{M}_{\sigma, \sigma \to \varphi, \varphi}
= 2 g_1 (g_1 + g_2) s^{\lambda^{[3]} - 1} \quad,
\eeq
\beq
\mathcal{M}_{\varphi, \varphi \to \sigma, \sigma}
- 2 \mathcal{M}_{\varphi, \sigma \to \sigma, \varphi}
+ \mathcal{M}_{\sigma, \sigma \to \varphi, \varphi}
= 2 g_1 (g_2 - g_1) s^{\lambda^{[4]} - 1} \quad,
\eeq
\beq
F_1 g_1^2 \mathcal{M}_{\varphi, \varphi \to \varphi, \varphi}
+ (F_2 g_2^2 - F_1 g_1^2)
\mathcal{M}_{\varphi, \sigma \to \varphi, \sigma}
- F_2 g_1^2 \mathcal{M}_{\sigma, \sigma \to \sigma, \sigma}
= 0 \quad,
\eeq
\beq\label{4d}
\mathcal{M}_{\varphi, \varphi \to \sigma, \sigma}
- \mathcal{M}_{\sigma, \sigma \to \varphi, \varphi} = 0 \quad,
\eeq
where, e.~g., $\mathcal{M}_{\varphi, \sigma \to \varphi, \sigma}$ is
short for 
$\mathcal{M}((\vec{p}_1, \varphi), (\vec{p}_2, \sigma) 
\to (\vec{p}_3, \varphi), (\vec{p}_4, \sigma))$.

Of course, these relations are only valid in the Regge limit 
$s \to \infty$, $t$ fixed, and apply only to Regge behaviour, so 
that a vanishing combination of amplitudes means that this combination 
does not exhibit Regge behaviour.

In the example, eigenvalues $\lambda^{[i]}$ can vanish only for
$g_2 = \pm g_1$. If, say, $g_2 = g_1$ (i.~e.\ 
$\lambda^{[1]} = g_1^2 (F_1 + F_2)$, $\lambda^{[2]} = 0$,
$\lambda^{[3]} = 2 g_1^2 F_3$, $\lambda^{[4]} = 0$), the relations
(\ref{4c}) -- (\ref{4d}) simplify to
\begin{eqnarray*}
\mathcal{M}_{\varphi, \varphi \to \varphi, \varphi}
= \mathcal{M}_{\sigma, \sigma \to \sigma, \sigma}
= \mathcal{M}_{\varphi, \sigma \to \varphi, \sigma}
& = & g_1^2 s^{\lambda^{[1]} - 1} \quad,\quad \\
\mathcal{M}_{\varphi, \varphi \to \sigma, \sigma}
= \mathcal{M}_{\sigma, \sigma \to \varphi, \varphi}
= \mathcal{M}_{\varphi, \sigma \to \sigma, \varphi}
& = & g_1^2 s^{\lambda^{[3]} - 1} \quad.
\end{eqnarray*}

\section{Concluding Remarks}

Our results obtained in the $s$-channel within the collinear
asymptotic approach agree with the $t$-channel Bethe-Salpeter
treatment. This puts on firm ground our interpretation of Regge
behaviour as a consequence of collinear three-particle dynamics
in the $s$-channel. It opens the way for the application of the
collinear asymptotic Hamiltonian in a time-ordered exponential
approximation to realistic theories. Especially, we have in mind
effective QCD-inspired hadron interactions \cite{ba} at large 
distances, and on the other hand jet production in the deep inelastic 
region.

\end{document}